\documentclass[a4paper,12pt]{article}
\usepackage[english]{babel}
\usepackage{graphicx}
\usepackage{cite}
\begin{document}

\begin{center}
\huge \bf Electron transport in armchair \\single-wall carbon
nanotubes
\end{center}

\begin{center}
\Large D.V. Pozdnyakov{\tiny{ }}\footnote{pozdnyakov@bsu.by}, V.O.
Galenchik, \\F.F. Komarov and V.M. Borzdov
\end{center}

\begin{center}
\footnotesize \it Radiophysics and Electronics department, Belarus
State University, \\Nezavisimosty av. 4, 220050 Minsk, Belarus
\end{center}

\bigskip

{\small \noindent {\bf Abstract}

\bigskip

The rates of electron scattering via phonons in the armchair
single-wall carbon nanotubes were calculated by using the improved
scattering theory within the tight-binding approximation. Therefore,
the problem connected with the discrepancy of the scattering rates
calculated in the framework of the classical scattering theory and
ones predicted by experimental data was clarified. Then these
results were used for the solving of the kinetic Boltzmann equation
to describe electron transport properties of the nanotubes. The
equation was solved numerically by using both the finite difference
approach and the Monte Carlo simulation procedure.}

\bigskip

\bigskip



\bigskip

\bigskip

{\bf 1. Introduction}

At present the interest to the study of carbon nanotubes increases
dramatically \cite{Dresselhaus1, Jishi2, Ando3, Hertel4, Sedeki5,
Yao6, Dresselhaus7, Taylor8, Lemay9, Klesse10, Xiao11, Wind12,
Heinze13, Javey14, Rotkin15, Dresselhaus16, Dresselhaus17, Odom18,
Odom19, Pozdnyakov, Park, Mann, Lazzeri}. First of all it is caused
by prompt development of nanoelectronics technologies. Due to these
achievements it is possible to fabricate the pure carbon nanotubes
with high-quality structure \cite{Odom18, Odom19}. The prospect of
their future application as active and passive elements of
nanoelectronics is out of doubts today \cite{Yao6, Taylor8, Xiao11,
Wind12, Heinze13, Javey14, Rotkin15, Dresselhaus16, Dresselhaus17}.
Moreover, it is necessary to note that the theoretical study of
carbon nanotubes presently has much to be desired. It concerns both
the descriptions of the physical properties of nanotubes, and the
descriptions of the charge carrier transport in such structures. In
our opinion, this problem is connected on the one hand by that the
nanotubes are specific objects of quantum nature. That is why the
study of them is rather complicated task. And on the other hand
these structures have been investigated intensively only one decade.

One of the important areas in investigations of the nanotubes is the
study of their electrophysical properties, especially the properties
of the armchair single-wall nanotubes. It is known that such
nanotubes can be reproduced in the best way, and their
electrophysical properties are practically independent on their
diameter \cite{Javey14, Rotkin15, Park}. This cannot be said about
the nanotubes with another chiralities. So, for example, zigzag type
nanotubes can have either metallic or semi-conducting properties
depending on their diameter \cite{Dresselhaus1}. That is why we will
consider armchair nanotubes only. Therefore, to avoid the account of
the edge effects let us consider rather long structures ($L > 20$
nm), i.e. we will consider the armchair nanotubes with the length
greater then electron mean free path \cite{Yao6, Javey14, Park}.

To describe the electron transport in such carbon nanotubes the
semiclassical approach and the kinetic Boltzmann equation for
one-dimensional electron gas can be used \cite{Yao6, Yamada20}. Both
the finite-difference method \cite{Yao6, Yamada20} and the Monte
Carlo imitative simulation \cite{Jacoboni21, Hess22, Ivaschenko23,
Borzdov24} can be used for solving of this equation. Nevertheless,
the correct use of these approaches is possible only when all the
dominating charge carrier scattering processes are described in the
most precision way. In single-wall nanotubes these scattering
processes are the phonon ones \cite{Ando3, Hertel4, Sedeki5, Yao6,
Klesse10, Javey14, Pozdnyakov, Park, Mann, Lazzeri}. At the same
time, as far as we know, this problem is very poorly studied yet.
Hitherto, in particular, the reason of a huge difference in
theoretical calculations and experimental definition of
electron-phonon scattering rates as well as the independence of the
latter from the diameter of nanotubes \cite{Javey14, Park} are not
cleared up. That is why the purposes of the present paper are to
calculate the rate of electron scattering via phonons in the
armchair single-wall nanotubes and to develop a model of electron
transport in such nanotubes based on the solution of the Boltzmann
transport equation.

\bigskip

{\bf 2. Theory}

Let us consider the phonon scattering in nanotubes at the electric
quantum limit, i.e. when the angular momentum of all electrons is
equal to zero \cite{Klesse10}. To meet this condition the nanotubes
with small diameter $d < 3$ nm are considered. Such nanotubes
correspond to $(n,n)$ armchair nanotubes with the chirality index $n
\le 20$ \cite{Dresselhaus1}. In this case the scattering rates can
be calculated by using the perturbation theory \cite{Davydov25}.
Then, taking into account the one-dimensional nature of the electron
gas in nanotubes, the following equations are valid
\cite{Pozdnyakov, Yamada20, Ivaschenko23, Calecki26, Borzdov27,
Borzdov28, Pozdnyakov29}

\begin{equation}
\label{eq1}
W_{\nu} ^{{\rm e/a}} (k) = {\frac{\omega _{\nu} ^{\rm
e/a}}{2} }\left( {N (\omega _{\nu} ^{\rm e/a}, T_{\nu}) +
{\frac{{1}}{{2}}}\pm {\frac{{1}}{{2}}}} \right){\left| {g_{\nu}
(k,k_{\nu} ^{{\rm e/a}} )} \right|}^{2}\left| {{\frac{{\partial
E}}{{\partial k}}}} \right|_{k = k_{\nu }^{{\rm e/a}}} ^{ - 1},
\end{equation}

\begin{equation}
\label{eq2}
N (\omega _{\nu} ^{\rm e/a}, T_{\nu}) = \left( {\exp
\left( {{\frac{{\hbar \omega _{\nu} ^{{\rm e/a}}}}{{k_{B} T_{\nu}} }
}} \right) - 1} \right)^{ - 1},
\end{equation}

\begin{equation}
\label{eq3}
k_{\nu} ^{{\rm e/a}} = k + q,
\end{equation}

\begin{equation}
\label{eq4}
E(k_{\nu} ^{{\rm e/a}} ) = E(k) \mp \hbar \omega _{\nu}
^{{\rm e/a}} (q),
\end{equation}

\noindent where $k$ is the electron wave vector; $q$ is the phonon
wave vector; $\hbar$ is the Plank constant; $E$ is the electron
energy relative to Fermi level $E_{F}$ which is taken as zero
($E_{F} = 0)$; $g$ is the matrix element of electron-phonon
coupling; $k_{B}$ is the Boltzmann constant; $\omega _{\nu}^{{\rm
e/a}}$ is the cyclic phonon frequency; $T_{\nu}$ is the phonon gas
temperature; $\nu$ is the index signing the phonon branch; ${\rm
e/a}$ are the indexes signing phonon emission or absorption.
Furthermore, it should be noted that in eq. (\ref{eq1}) the Pauli
principle is not taken into account \cite{Davydov25}. The latter can
be taken into consideration while the Boltzmann equation is being
solved or while the Monte Carlo simulation of the electron ensemble
drift checking whether the final state is free \cite{Jacoboni21,
Hess22, Ivaschenko23}. Moreover, the formula for the angular
momentums of electrons and phonons have not appeared in the system
of eqs. (\ref{eq3}) and (\ref{eq4}). This formula is not considered
by us because in the electric quantum limit the electrons will be
scattered by only the phonons with zero angular momentum due to the
angular momentum conservation. Therefore we will be taking into
account only such phonons.

The relation between the electron wave vector and the electron
energy is given by the known formula \cite{Jishi2, Satio33}

\begin{equation}
\label{eq6}
E_{1,2} (k) = \mp J_0 \left( 1 - 2 \cos{ \left( \frac{a
k}{2} \right) } \right),
\end{equation}

\noindent where $E_{1,2}$ is the electron energy in the band 1 and
2, respectively \cite{Jishi2, Hertel4, Rotkin15}; $a = 0.246$ nm is
the lattice constant \cite{Dresselhaus1, Sedeki5}; $J_{0} = 2.7$ eV
is the overlap integral \cite{Dresselhaus7, Lemay9}. The matrix
element of electron-phonon coupling can be calculated by using the
classical scattering theory within the tight-binding model
\cite{Jishi2, Barisic30, Pietronero31}, but only with some
improvements which we are dwelling on below. In this case the basic
equations for the calculation of the matrix element can be taken
from \cite{Jishi2, Pietronero31}

\begin{equation}
\label{eqgl}
\vert g_{l} \vert ^{2} = {\frac{{12q_{l}^{2}
J_{0}^{2}}} {{\mu _{l} \left( {\omega _{l}^{{\rm e/a}}}
\right)^{2}}}}\cos ^{2}\left( {{\frac{{\left( {k + k_{l}^{{\rm
e/a}}} \right)a}}{{4}}}} \right)\sin ^{2}\left( {{\frac{{\left( {k -
k_{l}^{{\rm e/a}}} \right)a}}{{4}}}} \right),
\end{equation}

\begin{equation}
\label{eqgt}
\vert g_{t} \vert ^{2} = {\frac{{4q_{t}^{2} J_{0}^{2}}}
{{\mu _{t} \left( {\omega _{t}^{{\rm e/a}}} \right)^{2}}}}\sin
^{2}\left( {{\frac{{\left( {k + k_{t}^{{\rm e/a}}} \right)a}}{{4}}}}
\right)\sin ^{2}\left( {{\frac{{\left( {k - k_{t}^{{\rm e/a}}}
\right)a}}{{4}}}} \right),
\end{equation}

\noindent where $l$ and $t$ are the indexes signing the longitudinal
and transverse phonon modes, respectively \cite{Jishi2}.

Having calculated the scattering rates of all possible phonon
scattering processes in the armchair nanotubes we have concluded
that the dominant scattering mechanisms are the longitudinal optical
(LO) and acoustic (LA) phonon backscattering processes as well as
the transverse acoustic (TA) phonon (twiston) backscattering. The
longitudinal phonon scattering is intraband with the transition from
one Dirac point to another \cite{Jishi2, Ando3, Hertel4, Klesse10,
Rotkin15}, whereas the TA scattering causes interband transition in
the vicinity of Dirac points. The other phonon scattering processes
can be neglected because their rates are very little in comparison
with the TA, LA and LO backscattering rates of active electrons, the
electrons, which are in a vicinity of the Dirac points ${\rm K}$ and
${\rm K'}$ close to the Fermi level. It should be noted that in
other studies only these dominant scattering mechanisms are taken
into account \cite{Hertel4, Yao6, Javey14, Pozdnyakov, Park, Mann,
Lazzeri}. Thus, the scattering rates in the armchair single-wall
nanotubes can be calculated according to eqs. (\ref{eq1}) --
(\ref{eqgt}) and the following formulae describing the dispersion
relations between the phonon wave vector $q$ and the phonon energy
$\hbar \omega$ for the phonons with zero angular momentum

\begin{equation}
\label{eqLA}
\hbar \omega _{{\rm LA}} (y) = {\left\{
{\begin{array}{l} {C_{4 -} ^{{\rm LA}} y^{4} - C_{3 -} ^{{\rm LA}}
y^{3} + C_{2 - }^{{\rm LA}} y^{2} + C_{1 -} ^{{\rm LA}} y,\quad y
\le 2 / 3,} \\  {C_{4 +} ^{{\rm LA}} y^{4} - C_{3 +} ^{{\rm LA}}
y^{3} + C_{2 + }^{{\rm LA}} y^{2} - C_{1 +} ^{{\rm LA}} y + C_{0 +}
^{{\rm LA}} ,\quad y > 2 / 3,}
\\ \end{array}} \right.}
\end{equation}

\begin{equation}
\label{eqLO}
\hbar \omega _{{\rm LO}} (y) = {\left\{
{\begin{array}{l} {C_{4 -} ^{{\rm LO}} y^{4} - C_{3 -} ^{{\rm LO}}
y^{3} + C_{2 - }^{{\rm LO}} y^{2} + C_{1 -} ^{{\rm LO}} y + C_{0 -}
^{{\rm LO}} ,\quad y \le 2 / 3,} \\ { - C_{3 +} ^{{\rm LO}} y^{3} +
C_{2 +} ^{{\rm LO}} y^{2} + C_{1 +} ^{{\rm LO}} y + C_{0 +} ^{{\rm
LO}} ,\quad y > 2 / 3,}
\\ \end{array}} \right.}
\end{equation}

\begin{equation}
\label{eqTA}
\hbar \omega _{{\rm TA}} (y) = {\left\{
{\begin{array}{l} {C_{1 -} ^{{\rm TA}} \sin \left( {1.880y +
2.411y^{5}} \right),\quad y \le 2 / 3,} \\ {C_{1 +} ^{{\rm TA}}
\left( {1 - 0.140\sin ^{2}\left( {3\pi y} \right)} \right)\cos
\left( {3\pi y} \right) + C_{0 +} ^{{\rm TA}} ,\quad y > 2 / 3,} \\
\end{array}} \right.}
\end{equation}

\begin{equation}
\label{eqy}
y = \vert ak \vert \left( {2\pi}  \right)^{ - 1}.
\end{equation}

\noindent There are $C_{4-}^{\rm LA} = 0.459$ eV, $C_{3-}^{\rm LA} =
0.645$ eV, $C_{2-}^{\rm LA} = 0.029$ eV, $C_{1-}^{\rm LA} = 0.359$
eV, $C_{4+}^{\rm LA} = 4.740$ eV, $C_{3+}^{\rm LA} = 17.28$ eV,
$C_{2+}^{\rm LA} = 23.33$ eV, $C_{1+}^{\rm LA} = 13.77$ eV,
$C_{0+}^{\rm LA} = 3.146$ eV, $C_{4-}^{\rm LO} = 0.918$ eV,
$C_{3-}^{\rm LO} = 1.100$ eV, $C_{2-}^{\rm LO} = 0.195$ eV,
$C_{1-}^{\rm LO} = 0.020$ eV, $C_{0-}^{\rm LO} = 0.196$ eV,
$C_{3+}^{\rm LO} = 0.069$ eV, $C_{2+}^{\rm LO} = 0.059$ eV,
$C_{1+}^{\rm LO} = 0.094$ eV, $C_{0+}^{\rm LO} = 0.083$ eV,
$C_{1-}^{\rm TA} = 0.124$ eV, $C_{1+}^{\rm TA} = 0.023$ eV and
$C_{0+}^{\rm TA} = 0.101$ eV in eqs. (\ref{eqLA}) -- (\ref{eqTA}).
These equalities have been found by using the analytical
approximation with high precision for the phonon-zone structure of
the armchair single-wall nanotubes. This zone structure is
determined by means of the zone folding method \cite{Satio33}
conformably to phonon-zone structure of graphene \cite{Wirtz32}. The
validity of such an approach is discussed in \cite{Lazzeri}.

Now let us consider the values ${\mu}_l$, ${\mu}_t$, $q_l$ and
$q_t$. Due to our investigations it is found out that there are
values ${\mu}_l = 2 m_C / a$ and ${\mu}_t = 4 m_C / a$ ($m_{C} = 2
\cdot 10^{ - 26}$ kg -- the mass of carbon atom) in ${\vert g_l
\vert}^2$ and ${\vert g_t \vert}^2$, respectively, instead of the
value of the armchair single-wall nanotube linear mass density
${\mu} = 4 n m_C / a$ \cite{Jishi2} (the first improvement of the
classical scattering theory). It is caused by equality of the carbon
atom oscillation phases on the whole circumference of the
cross-section perpendicular to the nanotube axis for the phonon
modes with zero angular momentum, i.e. by the synchronism of
oscillations of all the carbon atoms in a half of the primitive
cell. In that case it is possible to consider all the atoms in the
primitive cell like two, for the longitudinal modes, or four, for
the transverse ones, massive particles bounded with each other $2 n$
or $n$ times, respectively, as strong as two carbon atoms. The
discussed phenomenon clarifies the fact that the electron-phonon
scattering rates in the armchair single-wall nanotubes do not depend
on their diameters \cite{Javey14, Park} as the values ${\mu}_l$ and
${\mu}_t$ do not, too, in contrast to the value ${\mu}$.

As for the values $q_{l}$ and $q_{t}$, which are the deformation
constants in the directions along axis and circumference of the
nanotube, respectively, they can be calculated by using the
plane-wave approach. In contrast to the well-known Slater
local-orbital one giving the wrong results (see \cite{Hertel4}),
where the values $q_{l} = q_{t} = q_{0}$ are determined by means of
analysis of alteration of the overlap integral $J_{ov}$ due to the
distance alteration between two carbon atoms in graphene
\cite{Barisic30, Pietronero31}, in the plane-wave approach the
values $q_{l}$ and $q_{t}$ can be defined by the same way, but at
some different assumptions. The difference is that the Slater
local-orbital model assumes that the electrons are individual
particles belonging to individual atoms, whereas the plane-wave
model assumes that the electrons are de Broglie waves filling the
whole space in the crystal lattice of graphene. Moreover, it is
supposed that the phonon displaces only the ion but it does not
displace the electron orbit in crystal lattice within the Slater
local-orbital approach \cite{Barisic30, Pietronero31}, whereas
within the plane-wave one it is supposed that the phonon displaces
the whole carbon atom. In an approach like that, in fact, it is
taken into consideration the ``orbital relaxation'' (see
\cite{Hertel4}) when the relaxation time is assumed to be equal to
zero.

Then, using the results of \cite{Jishi2, Barisic30, Pietronero31,
Satio33}, the following formulae can be written for the unfolded
graphite sheet obtained from the armchair single-wall nanotube
within the plane-wave approach at the tight-binding approximation

\begin{equation}
\label{eqql}
q_{l} = {\frac{{1}}{{J_{0}}} }{\left. {{\frac{{\partial
J_{ov}}} {{\partial \vert {\rm \bf R}_{x} \vert}} }} \right|}_{R =
a} = {\frac{{1}}{{3}}}\left( {0 + {\frac{{\sqrt {3}}} {{2}}} +
{\frac{{\sqrt {3} }}{{2}}}} \right){\frac{{1}}{{J_{0}}} }{\left.
{{\frac{{\partial J_{ov} }}{{\partial R}} }} \right|}_{R = a},
\end{equation}

\begin{equation}
\label{eqqt}
q_{t} = {\frac{{1}}{{J_{0}}} }{\left. {{\frac{{\partial
J_{ov}}} {{\partial \vert {\rm \bf R}_{y} \vert}} }} \right|}_{R =
a} = {\frac{{1}}{{3}}}\left( {1 + {\frac{{1}}{{2}}} +
{\frac{{1}}{{2}}}} \right){\frac{{1}}{{J_{0}}} }{\left.
{{\frac{{\partial J_{ov}}} {{\partial R}} }} \right|}_{R = a},
\end{equation}

\begin{equation}
\label{eqJa}
{\left. {{\frac{{\partial J_{ov}}} {{\partial R}}}}
\right|}_{R = a} = {\left. {2 {\frac{{\partial}} {{\partial R}}}
{\left[ {\; {\int\limits_{\rm (primitive \; cell)} {V_{ov} (R) d{\rm
\bf r}}}} \right]}} \right|}_{R = a},
\end{equation}

\begin{equation}
\label{eqVov}
V_{ov} (R) = {\frac{{1}}{{(2\pi
)^{2}}}}{\int\limits_{\rm (Brillouin \; zone)} {E_{ov} (R,{\rm \bf
k}) d{\rm \bf k}}},
\end{equation}

\begin{equation}
\label{eqEov}
E_{ov} (R,{\rm \bf k}) = - J_{0} \sqrt {1 + 4\cos
^{2}\left( {Rk_{x} / 2} \right) + 4\cos \left( {Rk_{x} / 2}
\right)\cos \left( {\sqrt {3} Rk_{y} / 2} \right)}.
\end{equation}

\noindent Here ${\rm \bf r}$ is the two-dimensional radius-vector;
${\rm \bf k}$ is the two-dimensional wave vector; $x$ is the
coordinate along the nanotube axis $0x$; $y$ is the coordinate along
the nanotube circumference $0y$. In eq. (\ref{eqJa}) the coefficient
equal to 2 appeared because of there are two carbon atoms in the
primitive cell of graphene. According to eqs. (\ref{eqql}) --
(\ref{eqEov}) the value of $q_l$ is equal to $8.643 \; {\rm
nm}^{-1}$, and the value of $q_t$ is equal to $9.980 \; {\rm
nm}^{-1}$. Let us note that the values of $q_l$ and $q_t$ calculated
in the framework of the plane-wave approach (the second improvement
of the classical scattering theory) are less then the value of $q_0
= 25 \; {\rm nm}^{-1}$ calculated within the Slater local-orbital
approach \cite{Pietronero31}. Our results are in a good agreement
with the assumption of \cite{Hertel4} in which the values of $q_l =
8 \; {\rm nm}^{-1}$ and $q_t = 10 \; {\rm nm}^{-1}$ were considered
so as to secure the accordance of the theoretical results with the
experimental data.

For example, in fig.1 the functions of the electron-phonon
scattering rates versus the electron wave vector for the armchair
nanotubes calculated by using eqs. (\ref{eq1}) -- (\ref{eqy}) at
temperature $T_{\nu} = 290$ K are presented. It can be noted that
the LA and LO phonon scattering rates are of the same order whereas
the TA phonon scattering rate is about ten times less than those at
$T_{\nu} = 290$ K.

Let us write down the equations describing the state of the electron
gas in the armchair nanotube. In general, these equations can be
presented as \cite{Yao6, Pozdnyakov, Yamada20, Jacoboni21, Hess22,
Ivaschenko23, Borzdov24}

\begin{equation}
\label{eq19}
{\frac{{\partial f_{1,2}}} {{\partial t}}} +
{\frac{{\partial f_{1,2} }}{{\partial k}}}{\frac{{dk}}{{dt}}} +
{\frac{{\partial f_{1,2}}} {{\partial x}}}\left(
{{\frac{{dx}}{{dt}}}} \right)_{1,2} + \hat {I}f_{1,2} = 0,
\end{equation}

\begin{equation}
\label{eq20}
{\frac{{dk}}{{dt}}} = - {\frac{{e}}{{\hbar}} }F,
\end{equation}

\begin{equation}
\label{eq21}
\left( {{\frac{{dx}}{{dt}}}} \right)_{1,2} = u_{1,2} =
{\frac{{1}}{{\hbar }}}{\frac{{\partial E_{1,2}}} {{\partial k}}},
\end{equation}

\begin{equation}
\label{eq22}
n_{L} = {\frac{{1}}{{\pi}} }{\int\limits_{ - \pi / a}^{
+ \pi / a} {\left( {f_{1} + f_{2}}  \right)dk}},
\end{equation}

\begin{equation}
\label{eq23}
u_{dr} = {\frac{{1}}{{\pi n_{L}}} }{\int\limits_{ - \pi
/ a}^{ + \pi / a} {\left( {u_{1} f_{1} + u_{2} f_{2}} \right)dk}},
\end{equation}

\begin{equation}
\label{eq24}
J = - en_{L} u_{dr} .
\end{equation}

\noindent Here $t$ is the time; $f_{1,2} = f_{1,2} (k,x,t)$ is the
electron distribution function in the band 1 and 2, respectively;
$\hat {I}$ is the operator of the electron-phonon interaction with
account of the Pauli principle; $F = F(x,t)$ is the electric field
strength in the nanotube along $0x$; $u_{1,2} = u_{1,2} (k)$ is the
electron group velocity in the band 1 and 2, respectively; $n_{L} =
n_{L} (x,t)$ is the electron linear concentration; $u_{dr} = u_{dr}
(x,t)$ is the electron drift velocity; $J = J(x,t)$ is the electric
current in the nanotube; $e$ is the value of elementary charge.
According to \cite{Ivaschenko23} the operator $\hat {I}$ effects on
the electron distribution function as follows

\begin{equation}
\label{eq25}
\hat {I} f = \sum {\left( {Wf(1 - f') - W'f'(1 - f)}
\right)}.
\end{equation}

\noindent Here the sum is over all possible final states; $f'$ is
the electron distribution function in a final state; $W'$ is the
rate of scattering causing the return of electrons to the initial
state.

At the tight-binding approximation the eq. (\ref{eq21}) can be
written as (see the eq. (\ref{eq6}))

\begin{equation}
\label{eq26}
u_{1,2} = \mp {\frac{a J_0} {\hbar}} \sin{\left(
\frac{a k}{2} \right)}.
\end{equation}

So, to study the kinetic processes in the armchair single-wall
carbon nanotubes it is necessary to solve the kinetic Boltzmann
equation (\ref{eq19}) using the equations of motion (\ref{eq20}) and
(\ref{eq26}). Then the kinetic parameters $n_{L}$, $u_{dr}$ and $J$
can be calculated according to eqs. (\ref{eq22}) -- (\ref{eq24}).

\bigskip

{\bf 3. Results and discussion}

Let us consider electron transport in the infinite length armchair
nanotubes in uniform constant electric field. Such a transport is
practically equivalent to electron transport in rather a long
nanotube placed on the ideal metal contacts \cite{Yao6, Javey14,
Park}. Meanwhile, the nanotube can be considered as long if its
length is greater or at least the same order as the TA phonon
limited mean free path of active electrons. In this case the eqs.
(\ref{eq19}) and (\ref{eq22}) are reduced to

\begin{equation}
\label{eq27}
{\frac{e F} {\hbar}} {\frac{\partial f_{1,2} }
{\partial k}} - \hat{I} f_{1,2} = 0,
\end{equation}

\begin{equation}
\label{eq28}
n_{L} = {\frac{{2}}{{a}}}.
\end{equation}

\noindent The solving of eq. (\ref{eq27}) allows the kinetic
parameters to be calculated. Specifically it may be the electric
current in the nanotube.

In fig.2 and 3 some results of calculations of the electric current
in the armchair nanotubes at $T_{\rm TA} = T = 290$ K are presented.
Here $T$ is the temperature of the nanotube surroundings. These
results are obtained by using both the numerical solution of eq.
(\ref{eq27}) by the finite difference approximation and the
imitative Monte Carlo simulation of the electron ensemble drift
\cite{Borzdov34, Borzdov35, Borzdov36}. The results coincide and are
not resolved in the figure scale.

Moreover, the experimental data are presented in fig.2 and 3 along
with the theoretical results obtained by A. Javey et al.
\cite{Javey14}, who applied the Monte Carlo simulation procedure
with use of fitting parameters. These figures evidently demonstrate
a very good agreement between the theory and experiment. Especially
it concerns those results where the non-equilibrium LA and LO
phonons ($T_{\rm LA,LO} > T = 290$ K) \cite{Lazzeri} are taken into
account. Some discrepancy of the results in fig.3 at low fields is
explained by the TA phonon pinning effect at some conditions when
the TA phonon mode is partly suppressed due to an influence of the
substrate on oscillations of carbon atoms in the nanotube
\cite{Yao6}.

We did not consider the pinning effect, which is probably difficult
to describe theoretically, but we considered a most important effect
like heating of the LA and LO phonon gas, i.e. the non-equilibrium
LA and LO phonons \cite{Lazzeri}.  As a first order approximation
this effect can be described by the following obvious formula found
from the energy conversation law

\begin{equation}
\label{eq29}
N\left(\omega_{\rm K} ,T_{\rm LA,LO}\right) -
N\left(\omega_{\rm K} ,T\right) = \eta F J\left(F,T_{\rm
LA,LO}\right),
\end{equation}

\noindent where $\omega_{\rm K}$ is the cyclic frequency of the LA
and LO phonons at ${\rm K}$ point ($\hbar \omega_{\rm K} = 0.151$
eV) \cite{Wirtz32}; $\eta$ is the parameter which is determined by
the quality of the thermal contact between the nanotube and its
surroundings. The value of this parameter obviously depends on the
type of matter of the surroundings and the area of the contact as
well as another factors.

In our opinion, it is a very difficult task to calculate the value
of $\eta$ theoretically, but it is easy to find it from the
experimental data by using the fitting procedure to theoretical
results. So, we did like this. And the results presented in fig.2
and 3 have been obtained at such values of $\eta$ which give rise to
the best agreement between the theoretical results and experimental
data. In fig.4, in addition, the functions of the LA and LO phonon
gas temperature versus the electric field strength are given. These
functions have been calculated by means of eq. (\ref{eq29}) at the
best fitted parameter $\eta$. It is clear from this figure that the
non-equilibrium phonons cause the superheating of the LA and LO
phonon gas (see \cite{Lazzeri}).

We would like to add that the TA phonon pinning effect can be
described, in principle, by the same way like the superheating
effect of the LA and LO phonon gas in nanotubes. In other words it
is possible to consider the pinning effect as if the supercooling of
the TA phonon gas takes place ($T_{\rm TA} < T$). Then to achieve
the agreement between theory and experiment in fig.3, like it is in
fig.2, it must be guessed that the temperature of the TA phonon gas
$T_{\rm TA}$ is equal to 200 K ($T = 290$ K).

So, the validity of the improved scattering theory is verified by
the things discussed above and the results presented in fig.2 and 3.
Moreover, it is confirmed by the following reasons. The TA phonon
limited mean free path of active electrons $l_t$ as well as the LA
and LO phonon limited one $l_l$ calculated theoretically are in a
good accordance with the same values, respectively, found from the
experiments. For instance, the experiments, which the pining effect
is not observed in, predict that the value of $l_t$ is equal to $500
\pm 200$ nm at $T = 290$ K \cite{Javey14, Mann}. The improved
scattering theory predicts that the value of $l_t = u_F \left(
W_{\rm TA}^{\rm e}(k_F) + W_{\rm TA}^{\rm a}(k_F) \right)^{- 1}$
\cite{Park} is equal to 632 nm. Here $u_F$ and $k_F$ are the
absolute value of the group velocity and electron wave vector,
respectively, at the Fermi level in the armchair single-wall
nanotube. It also follows from the experiments that $l_l$ is equal
to 10\,--\,15 nm at very high electric fields ($\sim 10^7$ V/m)
\cite{Yao6, Javey14, Park}. The theory gives that the quantity of
$l_l = u_F \left( W_{\rm LA}^{\rm e}(k_F) + W_{\rm LA}^{\rm a}(k_F)
+ W_{\rm LO}^{\rm e}(k_F) + W_{\rm LO}^{\rm a}(k_F) \right)^{- 1}$
possesses the following values: $l_l (T_{\rm LA,LO} = 290 \; {\rm
K}) = 23.8$ nm, $l_l (T_{\rm LA,LO} = 1200 \; {\rm K}) = 15.0$ nm
and $l_l (T_{\rm LA,LO} = 1970 \; {\rm K}) = 10.0$ nm. According to
the results discussed above and taking into consideration the
superheating of the LA and LO phonon gas, one can draw a conclusion
that the improved scattering theory appropriately describes the real
scattering processes in the armchair single-wall carbon nanotubes.
Besides, it predicts that the temperature of the LA and LO phonon
gas is between 1000 K and 2000 K at very high electric fields. These
temperatures are much less and more reliable than those which were
guessed in \cite{Lazzeri} ($> 6000$ K).

Thus, in the present paper the rates of the electron scattering via
phonons in the armchair single-wall carbon nanotubes are calculated
by using the improved scattering theory. The model of charge carrier
transport in such nanotubes based on the solving of the Boltzmann
transport equation is developed. It is found out that the
discrepancy of theoretical calculations in the framework of the
classical scattering theory and experimental estimations of the
values of $l_l$ and $l_t$ has the following reasons: (i) the
features of carbon atom oscillations for the one-dimensional phonon
modes with zero angular momentum are not taken into account; (ii)
the excessive values of $q_l$ and $q_t$, obtained by Pietronero et
al. for graphene \cite{Pietronero31} by using the Slater
local-orbital approximation \cite{Barisic30}, are used; (iii) the
superheating of the phonon gas is not taken into consideration.
Everything noted above sets conditions for that there is a
discrepancy between the value of $l_l$ ($l_t$) calculated in the
framework of the classical scattering theory and the value of $l_l$
($l_t$) obtained experimentally. As an example, this discrepancy is
more than two times for $(10,10)$ nanotube. Therefore, to provide
the agreement of theory with experiment it is necessary to use the
values ${\mu}_l$ and ${\mu}_t$ instead of the value ${\mu}$ (the
first improvement of the classical scattering theory) as well as the
value of $q_l = 8.643 \; {\rm nm}^{-1}$ and $q_t = 9.980 \; {\rm
nm}^{-1}$ instead of the value of $q_0 = 25 \; {\rm nm}^{-1}$ (the
second improvement of the classical scattering theory) while
calculating the phonon scattering rates of electrons in armchair
single-wall carbon nanotubes. Moreover, it is also necessary to take
into account the non-equilibrium  LA and LO phonons at high electric
fields.

We think that we would succeed much more in the agreement of the
theoretical results with the experimental ones if we considered the
finite length nanotubes instead of the infinite ones. But in that
case the computational complexity would increase much more, too. So,
in conclusion we would like to say that our subsequent purpose is
the calculation of the electrical characteristics of the armchair
single-wall carbon nanotubes at various electric fields $F = F (t)$
in the framework of the developed scattering theory.

The authors gratefully wish to acknowledge Dr. D. Weiss  for his
careful examination of the paper.

\newpage
\noindent {\Large \bf Figure captions}

\bigskip

\noindent {\bf Fig.\,1.} LA, LO and TA phonon scattering rates in
the band 1 (a) and 2 (b) of the armchair single-wall carbon
nanotubes.

\bigskip

\noindent {\bf Fig.\,2.} The function of the electric current versus
the electric field strength in armchair single-wall carbon nanotubes
($\eta = 1.063 \cdot 10^{- 4} \; {\rm W}^{- 1} {\rm m}$).

\noindent The solid curve is the experimental data from
\cite{Javey14} ($L = 700$ nm).

\noindent The circles are the theoretical results from
\cite{Javey14} ($L = 700$ nm).

\noindent The points are our theoretical results at $T_{\rm LA,LO} =
290$ K ($L = \infty$).

\noindent The squares are our theoretical results in which it is
taken into account the non-equilibrium LA and LO phonons ($L =
\infty$).

\bigskip

\noindent {\bf Fig.\,3.} The function of the electric current versus
the electric field strength in armchair single-wall carbon nanotubes
($\eta = 6.737 \cdot 10^{- 4} \; {\rm W}^{- 1} {\rm m}$).

\noindent The solid curve is the experimental data from \cite{Park}
($L = 1000$ nm).

\noindent The points are our theoretical results at $T_{\rm LA,LO} =
290$ K ($L = \infty$).

\noindent The squares are our theoretical results in which it is
taken into account the non-equilibrium LA and LO phonons ($L =
\infty$).

\bigskip

\noindent {\bf Fig.\,4.} The function of the temperature of the LA
and LO phonon gas versus the electric field strength in the infinite
length armchair single-wall carbon nanotubes at $\eta = 1.063 \cdot
10^{- 4} \; {\rm W}^{- 1} {\rm m}$ (solid curve) and $\eta = 6.737
\cdot 10^{- 4} \; {\rm W}^{- 1} {\rm m}$ (dashed curve).

\end{document}